\documentclass[showpacs,prl,twocolumn,floatfix,reprint]{revtex4-2}
\usepackage{amsmath,graphicx,amssymb,hyperref}
\usepackage{tabularx,array,color}

\begin{document}

\title{Evolution of road infrastructures in large urban areas}

\author{Erwan Taillanter}
\email{erwan.taillanter@ipht.fr}
\affiliation{Universit\'e Paris-Saclay, CNRS, CEA, Institut de Physique Th\'{e}orique, 91191, 
Gif-sur-Yvette, France}

\author{Marc Barthelemy}
\email{marc.barthelemy@ipht.fr}
\affiliation{Universit\'e Paris-Saclay, CNRS, CEA, Institut de Physique Th\'{e}orique, 91191, 
	Gif-sur-Yvette, France}
\affiliation{Centre d'Analyse et de Math\'ematique Sociales (CNRS/EHESS) 54 Avenue de Raspail, 75006 Paris, France}

\begin{abstract}

Most cities in the US and in the world were organized around car traffic. In particular, large structures such as urban freeways or ring roads were built for reducing car traffic congestion. With the evolution of public transportation, working conditions, the future of these structures and  the organization of large urban areas is uncertain. Here, we analyze empirical data for US cities and show that they display two transitions at different thresholds. For the first threshold of order $T_c^{FW}\sim 10^4$ commuters, we observe the emergence of a urban freeway. The second  threshold is larger and of the order  $T_c^{RR}\sim 10^5$ commuters above which a ring road emerges. In order to understand these empirical results, we propose a simple model based on a cost-benefit analysis which relies on the balance between construction and maintenance costs of infrastructures and the trip duration decrease (including the effect of congestion). This model predicts indeed such transitions and allows us to compute explicitly the commuter's thresholds in terms of critical parameters such as the average value of time, average capacity of roads, typical construction cost, etc. Furthermore, this analysis allows us to discuss possible scenarios for the future evolution of these structures. In particular, we show that in many cases it is beneficial to remove urban freeways due to their large social cost (that  includes pollution, health cost, etc). This type of information is particularly useful at a time when many cities must confront with the dilemma of renovating these aging structures or converting them into another use. 
  
\end{abstract}


\maketitle

\begin{figure}[ht!]
	\includegraphics[width=0.5\textwidth]{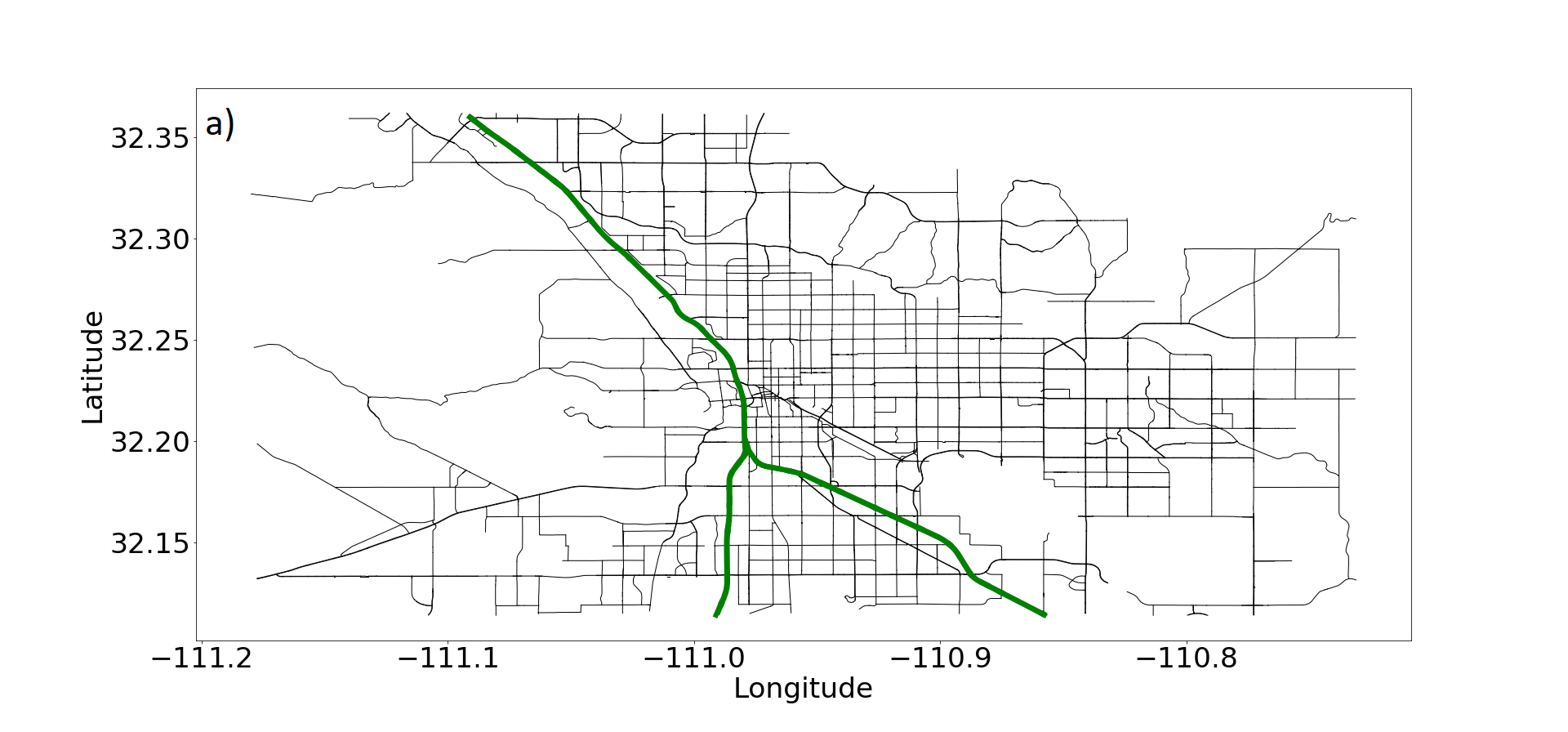}
	\includegraphics[width=0.5\textwidth]{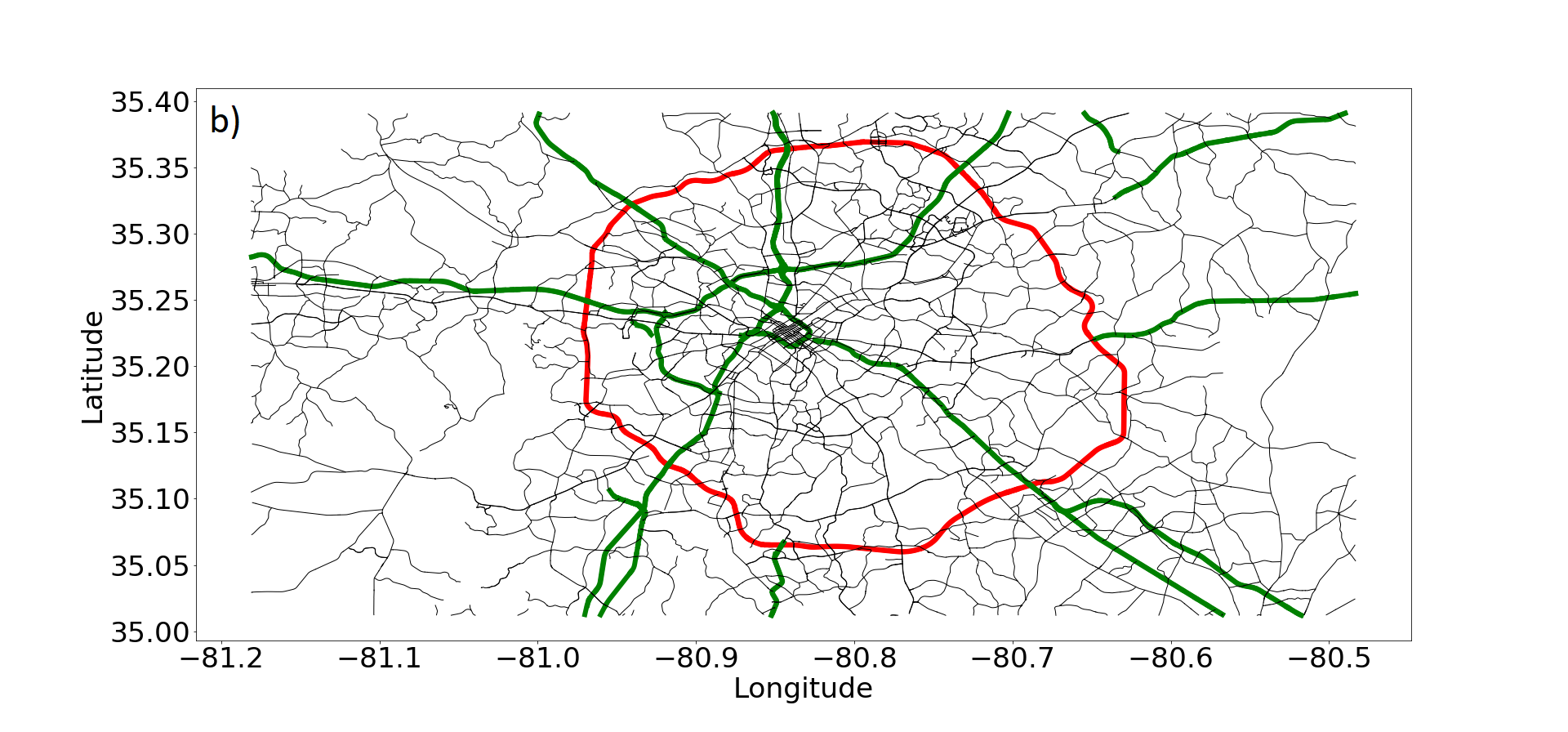}
	\caption{(a) Example of Tucson (AZ, USA, $540,000$ inhabitants in 2020), a city with two urban freeways (green) (b) Example of Charlotte (NC, USA, $860,000$ inhabitants in 2020), a city with a ringroad (red) and several urban freeways (green) }
	\label{fig:examples}
\end{figure}

In many cities around the world, a reflection about the allocation of space to cars has started and is now crucial. The car-centered paradigm which has shaped cities in the last century starts to be criticized, as its impact on the environment \cite{Brown:2009} \cite{Pojani:2020} \cite{vpti environmental cost land use}, social segregation \cite{Unfair repartition of land use} and on the overall well-being of the population \cite{Damages air pollution US} \cite{Impact noise} has now been extensively discussed.  In the US, pre-nineteenth century cities developed essentially following a grid pattern, together with radial Boulevards of larger sizes \cite{Conzen:2013}. With the generalization of cars in the first half of the 20th century, this infrastructure proved insufficient and road congestion became an issue in US-American cities as early as in the 1920s and 1930s \cite{Brown:2009},\cite{Pojani:2020},\cite{Brown:2002}. This issue was tackled by introducing traffic laws and traffic lights, and also by transforming the grid network of cities, adding a layer of expressways going right through their center (see an example in Fig.~\ref{fig:examples}a). 

The idea of urban expressways appeared as early as 1923 in New York \cite{Brown:2009},\cite{Brown:2002}. It was thought that by proposing a high capacity, high speed road, with limited junctions and no obstacles, the problem of congestion would be definitively solved. The first urban freeways were built in the 1930s, but it was during the 1950s and 1960s that they became a trademark of modernity in almost all but the smallest of US cities \cite{Brown:2009},\cite{Pojani:2020},\cite{Brown:2002}. The Federal Highway Act provided the large funding required to build new infrastructure and these funds were attributed prioritarily to projects promoting high car capacities, rather than multi-modal development or a good integration into the cities fabric, as investments were expected to be refunded by taxes on fuel \cite{Pojani:2020}. The promised reduction of congestion lasted for a few years or decades only, as the number of cars continued to increase. In addition, these high speed expressways triggered an exodus of the middle and upper classes as well as businesses themselves from the city centers to the suburbs, increasing the total distance traveled and in turn the congestion \cite{Hymel:2010},\cite{Goodwin:1996} \cite{Baum-Snow:2007}. To accommodate for the growing traffic demand between suburbs and to reduce congestion in city centers, it was thus decided to build new highways around the city center, usually referred to as `ring roads' (also known as `circular roads', `loops', `beltways' or `beltlines', see Fig.~\ref{fig:examples}b for an example). With a few early exception, most ring roads in the US were built in the late 1960s and onwards \cite{Brown:2009}, while in the rest of the world many ring roads were still being built after the turn of millennium (e.g. Beijing). Most of these infrastructures in the US are now over 60 years old and require substantial investments to be kept safe. The coming years thus appear to be the right time to think about sensible ways of using this budget to maximize the well-being of the population and change the paradigm of transport planning.

The emergence in urban areas of large infrastructures such as urban freeways or ring roads is thus a common fact and it is natural to study and understand the conditions under which this structure appears in cities. Although there were many studies, and for a long time \cite{Haggett:1969}, about the structure of street networks \cite{Jiang:2004,Marshall:2004,Buhl:2006,Crucitti:2006,Cardillo:2006,Lammer:2006,Xie:2007,Strano:2013,Boeing:2019}, and their evolution \cite{Samaniego:2008,Strano:2012,Barthelemy:2013,Xie:2011,kirkley2018,barthelemy2022}, few quantitative discussions adressed this problem of the emergence of large infrastructure. Some studies \cite{Ashton:2005,Jarrett:2006,Lion:2017} considered a ring and a central hub geometry and showed that the congestion effect at a central hub could be so large that avoiding the center is beneficial. Here in contrast, we discuss a topological transition where a new element appears in the geometry of the system. It is therefore much closer to problems discussed in network design in location science \cite{Laporte:2019,Aldous:2019}. In particular, it was shown in \cite{Aldous:2019} that when the total length available for constructing a transportation network is small, most of the resources go to the construction of radial branches and when it grows it becomes at a certain point beneficial to construct a ring.

In this article, we first perform an empirical study on road infrastructures in US cities and find population thresholds for the appearance of urban highways and ring roads. In order to understand the drivers of these transitions, we propose an analysis based on cost-benefit considerations where we take into account construction cost on one hand, and benefits in terms of time spent in traffic on the other. Under these assumptions, which mimic the car-centered paradigm of the 20th century, we show that there at two thresholds for population which agree with the observed historical evolution of urban road infrastructures. In the last part, we discuss the impact of urban highway removals. We show that by shifting the priorities and in particular by taking into account the environmental and social cost of urban highways, the optimum can shift towards their removal.

\section*{Empirical Study: Emergence of urban freeways and ring roads}

In this study, we focus on US cities and considered all micropolitan and metropolitan areas \cite{census bureau} and their population \cite{USpopu}, as defined by the US Census Bureau. The homogeneity of such a dataset is the main reason for this choice (more details on the data and this choice can be found in the Supplementary Information, along with other methodological details). 

The decision to build a urban freeway (or a ring road) was taken in response to the traffic demand. The relevant parameter here is thus the number of commuters at the time of construction of the road. One could argue that the area of the city could play a role in the shape of the infrastructure. The population density however displays very little variation among US cities, with a value of $\approx 10^3$ inhabitants/km$^2$ (see the SI for more details), and this seems to point towards the population (or the traffic demand) as the main control parameter of the problem. Considering the fact that most urban freeways and ringroads were built in the 1960s following the Federal Highway Act, we use the traffic demand in 1960 as the control parameter. We do only have the number of car commuters in 1960 for the biggest US-cities and we thus used these cities to extrapolate that the number of car-commuters $T_{1960}$ in each city of population $P_{2020}$ varies as $T_{1960}= rP_{2020}$ where $r\approx 0.14$ (see SI for details). 

\begin{figure}[ht!]
	\includegraphics[width=0.5\textwidth]{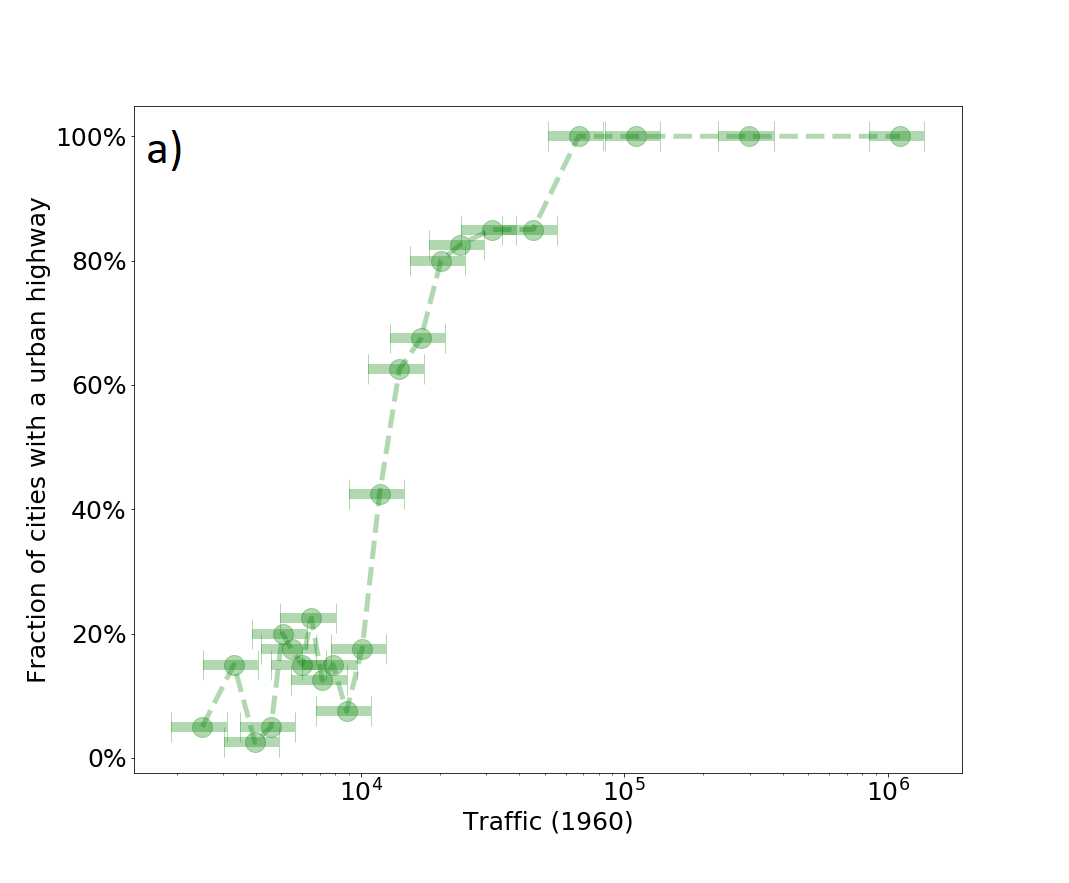}
	\includegraphics[width=0.46\textwidth]{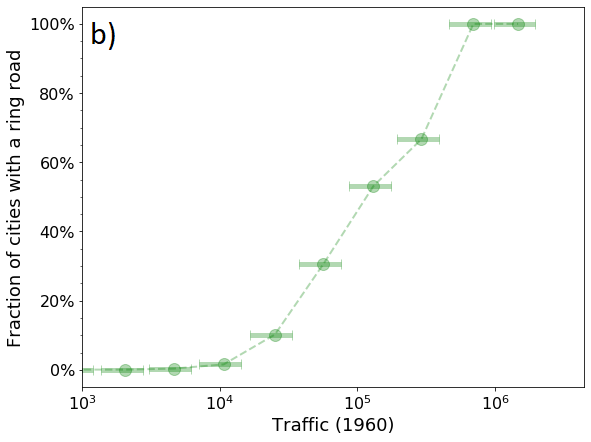}
	\caption{(a) Fraction of US cities with a urban freeway for a given number of commuters at the time of construction. Urban freeways appeared in cities with more than $10^4$ commuters. (b) Fraction of US cities with a ring road for a given number of commuters at the time of construction. Ring roads appeared in cities with more than $10^5$ commuters. The error bars correspond to the uncertainties on the number of commuters in 1960 (see details in the Methods section).}
	\label{fig:histos}
\end{figure}

Using Google Maps and the StreetView functionality, we assess whether a city has a urban freeway or not. We then sort cities by increasing traffic demand and compute the fraction of cities with a freeway in each bin of 40 cities (results are shown in Fig.~\ref{fig:histos}a).
We observe a sharp transition between small cities without urban freeways and larger cities with a urban freeway, for a critical traffic demand $T_c\approx 10^4$. In terms of current population, we find a transition for $P_c \approx 10^5$ inhabitants. In particular, only three urban areas  with population larger than $300,000$ have no urban freeway, namely the Metropolitan urban areas of  Evansville (IN, $311,552$ inhabitants), South Bend (IN, $319,224$ inhabitants) and the biggest urban area without a freeway, Rockford (IL, $349431$ inhabitants). Apart from these examples, this infrastructure, despite its cost and its impact on city planning, is completely generalized in the United States.

For ring roads, using data from \cite{wiki} we apply the same method (details in the Methods section) and compute the fraction of cities with a ring road in each bin of 40 cities (see Fig.~\ref{fig:histos}b). Here also, we observe a transition (albeit not as sharp as the previous one) between small and medium cities which generally don't have a ringroad and larger cities with a ring road. This transition occurs for traffic demand $T_c \approx 10^5$, which translates into a 2020 population of $\approx 10^6$ inhabitants. Not surprisingly, we find that all major cities with population $>10^7$ inhabitants have a ring road.


\section*{A cost-benefit analysis of these transitions}

We have empirically established that the presence of urban freeways and ringroads in US cities is almost entirely determined by the traffic demand and thus indirectly by the population. We found that there is a first critical demand for the emergence of urban freeways, followed by a second critical traffic demand for ring roads. In this section, we consider a simple toy-model based on cost-benefit considerations, to understand the evolution of infrastructures in terms of transitions between different optimal states.

\subsection*{The model and the cost-benefit framework}

\begin{figure}[ht]
	\includegraphics[width=0.5\textwidth]{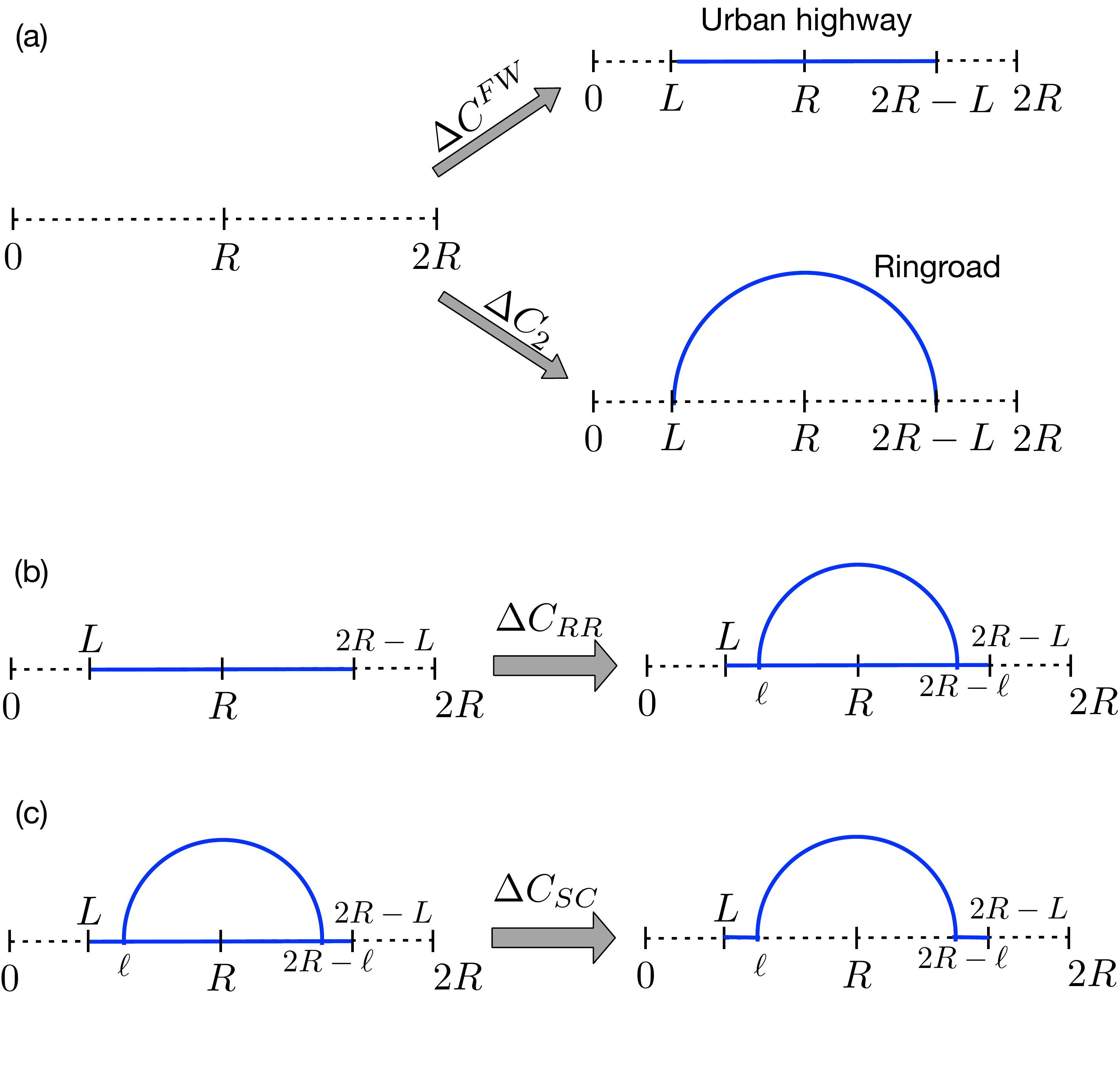}
	\caption{One-dimensional model of a city of size $2R$. (a) We first consider the case where a city has two possible choices: either to build a urban freeway of size $2(R-L)$ or a ring road of length $2g(R-L)$. (b) We then consider the possibility of adding a ring road between points $\ell$ and $2R-\ell$ to a city which has already a freeway. The corresponding cost difference is $\Delta C^{RR}$. (c) Finally, we discuss the possibility of removing the urban freeway between $\ell$ and $2R-\ell$. The corresponding cost is $\Delta C^{SC}$.}
	\label{fig:toymodel}
\end{figure}

We consider a one-dimensional city of size $2R$ depicted in Fig.~\ref{fig:toymodel} and where the density of car-commuters is uniform and given by $\rho = \frac{T}{2R}$.
$T$ describes the number of commuters who travel to work by car each day. All $T$ commuters are assumed to complete their trip during the morning (evening) rush hours, which we suppose are $h = 2$ hours long. Thus at any given time of the rush-hours, the density of people leaving their house is $\frac{T}{2 h R}$. The free-flow velocity in this city is $v_0$ and the free-flow velocity on the freeway or the ring road is assumed to be faster $v>v_0$. 

We will consider commuting in the morning in this city and for the sake of simplicity, we will consider the left part of the city only. We will assume that there are two types ($1$ and $2$) of commuters with different destinations. The type $1$ comprises a fraction $\alpha$ of individuals that commute to the center $R$ of the city, while the other type ($2$) of commuters (with fraction $\beta$ such that $\alpha+\beta=1$) commute to the point $2R$, essentially accounting for individuals traveling from suburb to suburb (for definitions and notations see table 1).


We illustrate the cost-benefit analysis used in this paper on the ring road case, but this discussion can be extended to any modification of the infrastructure. We denote by $\tau_{1(2)}$ the total time (summed over all commuters) to go to $R$ ($2R$) when there is no ringroad. In the presence of the ring road (with free velocity $v>v_0$), the corresponding times are denoted by $\tau_{1(2)}'$. The cost-benefit analysis relies on the cost variation (computed over typically a year) and defined as
\begin{align}
	\Delta C=2 g(R-\ell)\epsilon+2NV(\Delta\tau_1+\Delta\tau_2)
	\label{eq:Z}
\end{align}
where the first term denotes the construction (and maintenance) cost of the ringroad ($\epsilon$ is the cost per unit length of the ring road, $g$ is a geometric factor accounting for the shape of the ringroad and allowing us to write its length as $2g(R-\ell)$). The second term is the yearly benefit due to the ringroad, where $V$ is the average value of time of users and N the number of hours on which we perform our study. The factor $2$ accounts for indivuals commuting from both the left and right part of the city.
The quantity $\Delta\tau_i=\tau_i'-\tau_i$ is the time saved (per commute) by commuters of type $i$ thanks to the new infrastructure. The optimal size of the ring road is determined by the minimization condition of this cost-benefit quantity: $\min_{\ell}\Delta C$.  If the cost $\epsilon$ is very large, the minimization of $\Delta C$ will essentially be equivalent to take a minimal ring road size (at the limit, no ring road at all). In contrast, when the value of time $V$ is very large, the priority in this city is to minimize the commuting time, whatever the cost of the ringroad. In general, we can then expect that for a population large enough, building a ringroad to split the traffic between the two types of commuters will be beneficial.

In Eq.~\ref{eq:Z}, we only considered the cost of construction and the value of time spent commuting, but ignored other external cost of freeways. This is a strong assumption, which we believe to be close to the considerations at the time of construction of most freeways and ring roads under the Federal freeway Act, where the main concern was to maximize traffic capacity, regardless of urban integration \cite{Brown:2002}, \cite{Baum-Snow:2007}. In the last section of this article, we will discuss the effect on the optimal situation when one takes further costs into consideration.

\subsection*{Congestion for non uniform flows}

In order to compute the trip durations we take into account congestion thanks to the Bureau of Public road function \cite{Branston:1976}. More precisely, if there is traffic $Q$ on a road segment of length $d$ and of capacity $Q_c$, the trip duration is given by \cite{Branston:1976}
\begin{align}
	\tau=\frac{d}{v_0}\left[1+\left(\frac{Q}{Q_c}\right)^\mu\right]
\end{align}
where $v_0$ is the free-flow velocity on this road segment and $\mu$
an exponent that characterizes the sensitivity to congestion of this
road (typically $\mu\in [2,4]$ but we will use the simplifying
assumption of $\mu=1$ in order to get analytical results). 

Individuals are distributed in the city and have thus different starting points. This implies that the traffic flow will not be constant and we thus have to adapt the standard Bureau of Public Roads function \cite{Branston:1976} to this case. We first consider a road from $0$ to $R$ (with capacity $Q_c$ and free velocity $v_0$) and the commuting population is distributed according to a density $\rho(x)$. People leave their house at any possible time during a rush hour window of $h=2$ hours. All individuals commute to the point $R$ and we thus assume that the flow $Q(x)$ at point $x$ is equal to the number of people who left home upwards
\begin{align}
	Q(x)=\int_0^x\frac{\rho(y)}{h}\mathrm{d}y
\end{align}
For the elementary segment between $x$ and $x+\mathrm{d}x$, the infinitesimal
trip duration is according to the Bureau of Public road function
\begin{align}
	\mathrm{d}\tau(x)=\frac{\mathrm{d}x}{v_0}\left[1+\left(\frac{Q(x)}{Q_c}\right)^\mu\right]
\end{align}
and the total time to go from a point $a$ to the point $b$ on the road is 
\begin{align}
	\tau_0(a,b)=\int_a^b\frac{\mathrm{d}x}{v_0}
	\left[1+\left(\frac{Q(x)}{Q_c}\right)^\mu\right]
\end{align}
In the one dimensional problem the total time spent travelling by individuals located between $0$ and $R$ and going to $R$ is then 
\begin{align}
	\nonumber
	\overline{\tau}(R)&=\int_0^R\mathrm{d}x\rho(x)\tau_0(x,R)\\
	&=\int_0^R\mathrm{d}x\rho(x)\int_x^R\frac{\mathrm{d}x'}{v_0}
	\left[1+\left(\frac{Q(x')}{Q_c}\right)^\mu\right]
\end{align}
This calculation constitutes the basis of our cost-benefit analysis in more complicated cases with urban freeways and/or ring roads. Also, in the following we will use the value $\mu=1$ which will lead to linear terms of the form $\frac{T}{hQ_c}$ (instead of $\frac{T}{hQ_c}^\mu$) and will allow analytical calculations. Numerical calculations show however similar results for other values of $\mu$ (see figure SI4 for details). 

\subsection*{Possible scenarios: Urban freeway or ring road ?}

This cost-benefit framework allows us to discuss possible scenarios for 
the evolution of road infrastructures in urban areas. We first discuss here 
the case of smaller areas when two possibilities can be envisioned: either a urban freeway or a ring road. This alternative is schematically described in Fig.~\ref{fig:toymodel}(a). For the sake of simplicity, we assume here that both the urban freeway and the ring road connect the points $L$ and $2R-L$. The initial cost is given by
\begin{align}
	C^0=V(\tau_1+\tau_2)
	 = \frac{NVTR}{2v_0}\left[1+\frac{2}{3}\tilde{T}+2\beta(1+\beta\tilde{T})\right]
\end{align}
where $\tau_1$ and $\tau_2$ are the commuting times corresponding to $\alpha$ and $\beta$ commuters, respectively, and $\Tilde{T}=\frac{T}{2hQ_c}$($=\frac{cP}{2hQ_c}$), with $Q_c$ the capacity of each road, $T$ the total traffic demand, $h$ the duration of the rush hours, $P$ the population and $c$ the ratio between traffic demand and population.

\subsubsection*{Urban freeway case}

When we build a urban freeway (Fig.~\ref{fig:toymodel}a), the total cost is given by 
\begin{align}
	C^{FW}=2(R-L)\epsilon+2NV(\tau'_1+\tau'_2)
\end{align}
where the first term corresponds to construction and maintenance costs ($\epsilon$ is the cost per unit length) while the commuting times are modified due to the presence of the freeway and now read $\tau'_{1(2)}$. Following the strategy described above for non-uniform flows, we compute these different terms and we obtain for the total cost variation (see details in the SI)
\begin{align}
\begin{split}
	\Delta C^{FW} &=C^{FW}-C_0 \\
	 &=2R(1-y)\epsilon-\frac{NVTR}{2v_0}(1-\eta)\Big[1-y^2+\frac{2}{3}\tilde{T}(1-y^3)\\
	&+2\beta(1+\beta\tilde{T})(1-y)\Big]
\end{split}
\end{align}
where $y=\frac{L}{R}$ and $\eta=\frac{v_0}{v}$. The evolution of $\Delta C^{FW}$ with the population is illustrated in Fig.~\ref{fig:cost} and is first positive and gets negative for a critical population $T_c^{FW}$ above which building a freeway becomes cost effective. Keeping only the terms of highest order, the expression for this critical population is:
\begin{align}
T_c^{FW} \approx \sqrt{\frac{4}{(1+\beta^2)(1-\eta)}}\sqrt{\frac{\epsilon hv_0 Q_c}{NV}}
\end{align}
For typical values $v_0=20km/h$, $v=100km/h$ (which correspond to realistic values for urban environments with traffic lights, speed regulations, etc.), $V=20\$/hour$ and $N=730$ (i.e. considering the cost of congestion over a year, with each time morning and evening rush hours), $h=2$ hours, $\epsilon = 10^7 \$$/km, $Q_c=10^3 cars/hour$ and $\beta=0.5$, we find a critical demand $T_c^{FW}\approx 10^4$, consistent with our empirical results.

\subsubsection*{Ring road case}

If instead of building a freeway, we add a ring road (see Fig.~\ref{fig:toymodel}a) used by all of the type-2 commuters to travel from a suburb to another, the total cost variation is given by
\begin{align}
	\Delta C_2=2R(1-y)g\epsilon+2NV(\Delta\tau_1+\Delta\tau_2)
\end{align}
where the first term corresponds to the construction of the ring road ($g=\pi/2$ in the case of a half-circular ring road presented here, $y=\frac{L}{R}$) and the second term to the trip duration variations due to the presence of the ring road. Calculations are similar to the case described above (see SI for further details) and we find
\begin{align}
	\nonumber
	\Delta C_2&=2R(1-y)g\epsilon+\frac{NVTR}{2v_0}\Big[2\beta y(y-1)\\
	\nonumber
	&+\frac{2}{3}\tilde{T}\left[(1-y^3)(\alpha^2-1)+\beta^2(1-y)^3\right]\\
	&+2\beta(1+\beta\tilde{T})(1-y)(2\eta g-1)\Big]
\end{align}
The evolution of $\Delta C_2$ with the population is illustrated in Fig.~\ref{fig:cost}. We also find in this case a critical number of commuters $T_{c2}$ for which it becomes cost efficient to build a ring road around the city and which is given by
\begin{align}
	T_{c2} = \sqrt{\frac{2g}{\beta (1-\beta \eta g)}}\sqrt{\frac{hv_0 \epsilon Q_c}{NV}}
\end{align}
The condition of existence for this transition is to have $\beta \eta g<1$, which is typically fullfilled for realistical values ($\beta<1$ by definition, $g\approx \frac{\pi}{2}$ and the ratio of speeds between city and freeway is smaller than $\frac{1}{3}$). We note that, as expected, when $\beta\to 0$ the threshold $T_{c2}\to\infty$ since all commuters go to the center and won't need the ringroad. 

We note that this cost-benefit analysis allows us to express the population thresholds in terms of relevant parameters such as the construction cost per unit length, the velocity on the ringroad, the value of time, and the capacity of roads. Some part of these expressions could have been predicted using dimensional considerations but the square root behavior is not obvious and could not have been found by naive arguments. The population threshold thus increases as the square root of the capacity, which demonstrates the low impact of increasing capacity planning measures.

\subsubsection*{Urban freeways are more cost-efficient than ringroads}

Both $\Delta C^{FW}$ and $\Delta C_2$ show a transition when the population is increasing (see Fig.~\ref{fig:cost}), with critical traffic demands $T_c^{FW}$ and $T_{c2}$ respectively. Both critical demands share the factor $\sqrt{\frac{\epsilon h v_0 Q_c}{NV}}$, with a different prefactor that depends on details such as $\beta$, $\eta$ and $g$. Analyzing these expressions, we find that for realistic values of these parameters, we have $T_c^{FW}<T_{c2}$ regardless of $\beta$. This result shows that from a purely cost-benefit point of view, a urban freeway is preferable compared to a ring road and will thus appear as the first step in the evolution of road infrastructure in urban areas. This result is consistent with the historical evolution of road infrastructures in cities where urban freeways are indeed the first to appear and for a number of commuters above a certain threshold as we saw in the previous empirical section.  If a ring road appears, it might then be on top of this system comprising a urban freeway, a scenario that we will analyze in the next section.

\subsection*{Emergence of a ring road}

\begin{figure}[ht!]
\centering
\includegraphics[width=0.55\textwidth]{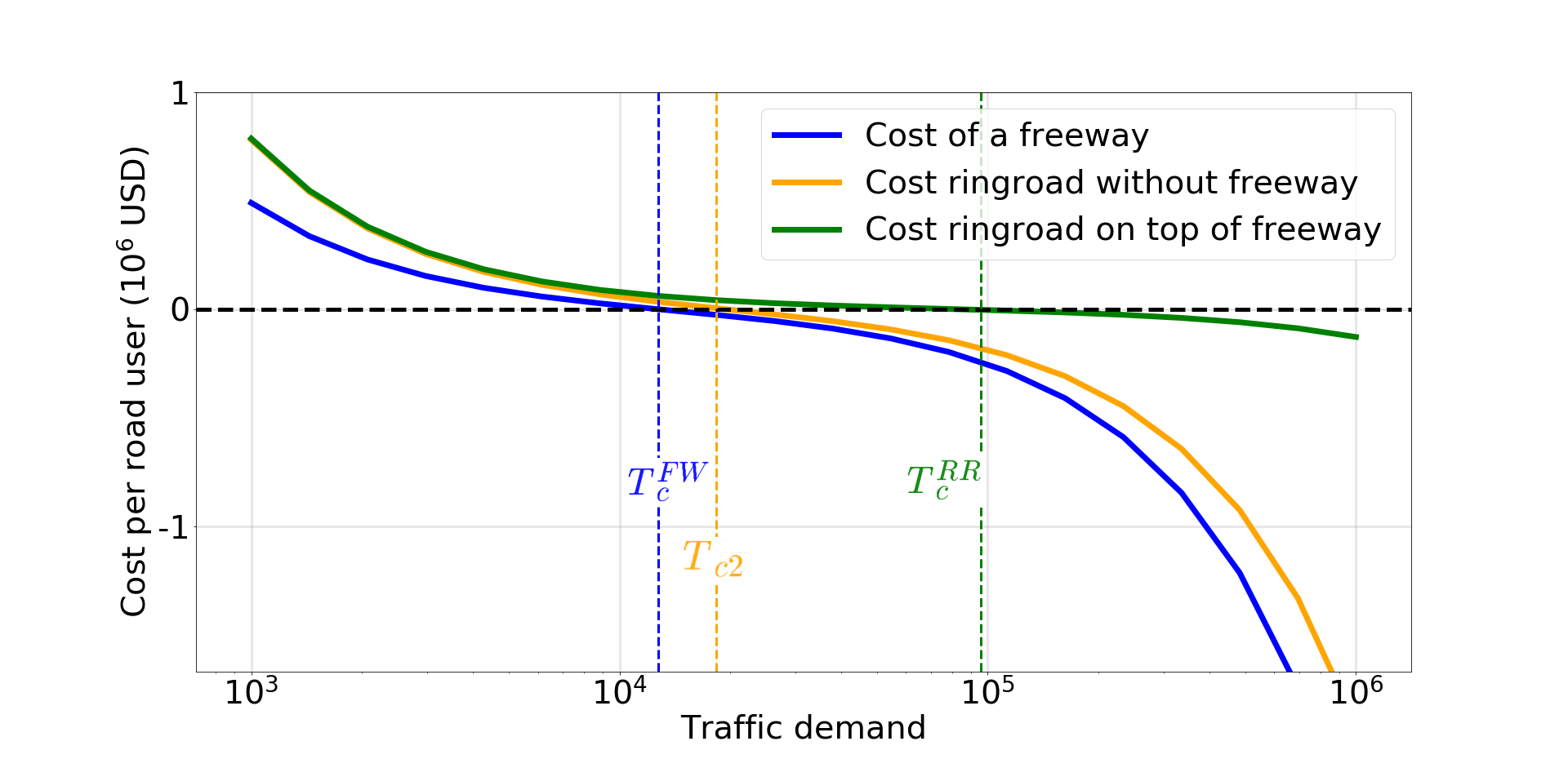}
\caption{Cost per road user for the construction of different infrastructures, as a function of the number of car commuters.}
\label{fig:cost}
\end{figure}

We now consider our previous system where a urban freeway of given size ($L$ is given) has been built. We add a ring road (see Fig.~\ref{fig:toymodel}(b)) connecting the points $\ell$ and $2R-\ell$ (the length of the ring road is $2g(R-\ell)$ where $g=\pi/2$ for a circular shape). The total cost difference reads (the details of the calculation can be found in the SI )
\begin{align}
	\Delta C^{RR}=2(R-\ell)g\epsilon+2NV(\Delta\tau_1+\Delta\tau_2)
\end{align}
The behavior of this quantity is shown in Fig.~\ref{fig:cost} and displays the existence of a critical population $T_c^{RR}$ above which adding a ringroad becomes the most cost efficient option and whose expression is
\begin{align}
	T_c^{RR}=\sqrt{\frac{4g}{\eta \beta (1-\beta g)}}\sqrt{\frac{\epsilon h v_0 Q_c}{NV}}
\end{align}
Using the numerical values previously discussed (see table \ref{table}), we find that $T_c^{RR}\sim 10^5$ whose order of magnitude is in agreement with our empirical observations. In the Fig.~\ref{fig:cost}, we also observe that the positive impact of a ringroad is greatly reduced when the city already has a freeway. In particular, this Fig.~\ref{fig:cost} is consistent with the empirical result that the transition for the appearance of freeways is relatively sharp at $\approx 10^4$ commuters, while the transition for ringroads is broader, in the range $10^4 - 10^6$ commuters.

The condition of existence for this transition is to have $\beta >0$ and $\beta < \frac{1}{g}$ and unlike the previous cases, this condition is not automatically fulfilled. Indeed, in the presence of a urban freeway, adding a ringroad is only beneficial when it allows to separate the two types of commuters $1$ and $2$. Rather counter-intuitively, for large values of $\beta$ (i.e. a lot of suburb to suburb commuters), building a ringroad actually is not beneficial. In order to understand this fact, we consider the extreme case $\beta=1$. Without the ring road, all commuters use the freeway through the city center. With the ring road, all commuters use the longer way around the city, resulting in a longer commute. It is only if a significant part $\alpha$ of commuters can benefit from the reduced traffic in the center that the ring road becomes an efficient solution. Similarly, if $\beta=0$, the addition of a ringroad is useless and $T_c^{RR}$ tends to infinity.

\section*{Social cost and freeway removal}

In large cities the existence of a urban freeway allows for a large velocity in the city center but also creates a number of problems such as pollution, noise, etc. Due to this discomfort - or `social cost' - created by these urban freeways, we observe a recent trend where cities decided to close these urban freeways and to use the space created for other purposes such as green spaces, etc. \cite{Brown:2009},\cite{Pojani:2020},\cite{Kraft-Klehm:2015}. With our framework we can analyze this phenomenon and in particular, we can provide an estimate of the social cost above which it becomes beneficial to close the urban freeway in the central area of the city.

In order to describe the problems triggered by the presence of the urban freeway, we assume that it comes with a social cost $\epsilon'$ per unit length (the maintenance cost per unit length of the freeway or ring road is still denoted by $\epsilon$). We analyze the cost difference between the case where there is both a urban freeway (size $y=L/R$) and a ring road (size $x=\ell/R$), and the case where we remove the  central freeway between $\ell$ and $2R-\ell$ (Fig.~\ref{fig:toymodel}(c)). The cost before the freeway removal is
\begin{align}
	\nonumber
	C&=[2(R-L)+2(R-\ell)g]\epsilon+NV(\tau_1+\tau_2)\\
	&+2(R-\ell)\epsilon'
\end{align}
and after the freeway removal (between $\ell$ and $2R-\ell$) social costs are removed and the trip durations are increased. The new cost then reads 
\begin{align}
	C'=[2(\ell-L)+2(R-\ell)]\epsilon+NV(\tau'_1+\tau'_2)
\end{align}
and the cost difference is given by 
\begin{align}
	\Delta C^{SC}=-2(R-\ell)(\epsilon'+\epsilon)+2NV(\Delta\tau_1+\Delta\tau_2)
\end{align}
If $\Delta C^{SC}<0$, the social cost is too high, and it is then beneficial to remove the urban freeway. The expression of $\Delta C^{SC}$ can be found in the SI. If we impose the condition that the freeway removal is beneficial regardless of the size of the ring road, we find that $\Delta C^{SC}<0$ for $\epsilon'>\epsilon'_c$ where
\begin{align}
	\epsilon'_c&=(1-\eta)\frac{NV\alpha^2}{4hv_0 Q_c}T^2-\epsilon
&= (1-\eta)\frac{NV\alpha^2}{4hv_0 Q_c} c^2P^2-\epsilon
\label{eq:epsilon'}
\end{align}
where the ratio $c$ between number of car-commuters and population is fixed for a given city and a given year, and shows little variation between cities (see Fig.~SI3 and \cite{Commuting 2013}). For 2020, we find $c \approx \frac{140 . 10^6}{330 . 10^6} \approx 40\%$  \cite{Commuting 2013}, \cite{pop:2020}.

\begin{figure}[ht!]
\centering
\includegraphics[width=0.5\textwidth]{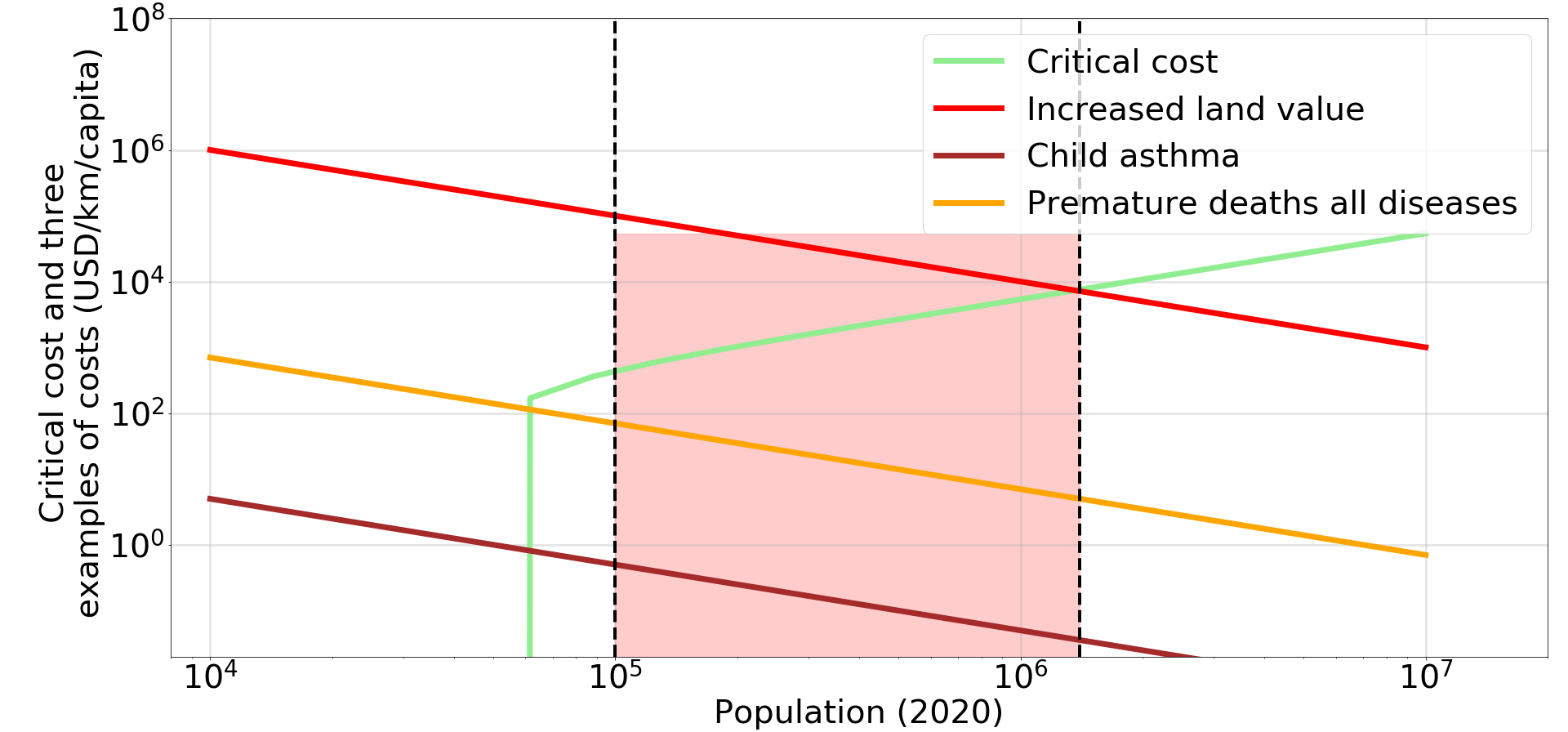}
\caption{Value per km of urban freeway for different external costs and critical value above which the freeway removal is economically sensible. We used values computed for the city of Seoul (South Korea) for estimating the land value cost (see text). The red surface (defined by $\epsilon'>\epsilon_c$ -the red line is above the critical green line)  shows cities which typically have a urban freeway (population $>100,000$) and would benefit to remove it (population $<1,400,000$).}
\label{fig:socialcost}
\end{figure}

We represent on the Fig.~\ref{fig:socialcost} this critical value per capita ($\epsilon'_c/P$) versus the population.

Below a critical value of the population, this threshold is zero, and above varies as Eq.~\ref{eq:epsilon'}. In order to estimate the social cost of a urban freeway, we can consider different impact of such an infrastructure. First, its impact on health such as child asthma or premature death due to air pollution. Evaluating the external costs associated to air pollution is an active field of research and it is difficult to get precise figures. There is however a consensus on the following points. First, air pollution increases the risk of lung-diseases (e.g. asthma, bronchiolitis), leading to medical cost and even premature deaths \cite{asthma incidence}. Air pollution is dramatically higher in the vicinity of major roadways and decays rapidly with distance (typically returns to baseline values within $\approx 500$m) \cite{near highway levels}. Finally, trees or vegetation have an extremely positive impact on the quality of the air close to roads \cite{Brugge:2015,Pugh:2012}. In order to have a quantitative estimate, we follow a canadian study that evaluates the cost of premature deaths caused by the Highway 401 in Toronto \cite{pollution:Toronto}. They estimated that this $\approx 50$km long highway reduces life expectancy in its vicinity corresponding to a total cost of $330$ millions USD  leading to $\approx 7$ millions USD per kilometer and the corresponding curve per capita is shown in Fig.~\ref{fig:socialcost}. Another example concerns traffic-induced asthma in children living close to highways. Using \cite{asthma incidence} \cite{near highway levels} \cite{children census} \cite{Population near highways}, we extrapolate that children living within $500$m of a highway have $15\%$ chances to develop asthma, as compared to the $10\%$ national baseline. The annual cost of one child having asthma is $\approx 4000\$$ \cite{asthma price}. We then extrapolate a cost per km associated to the presence of a highway of $\approx 40,000\$$/km/year shown per capita in Fig.~\ref{fig:socialcost} (see details in the SI). These are examples of cost associated to air pollution but there are several others which add each other to the social cost of a freeway \cite{Evaluation health impact}. 

Another important cost of urban freeways corresponds to the value of the land they occupy. This occupied surface in the city center could be exploited for residential or commercial buildings contributing to the cities GDP. In the case of the removal in 2005 of the Cheonggyecheon freeway in Seoul \cite{Cervero land capitalization}, one could observe an enhanced attractivity of the surroundings with an increase of land value of $3.10^3/m^2$ to $7.10^3/m^2$ for an area over a kilometer away from the freeway. Considering that the values for Seoul are similar to those in US cities (the GDP per capita is only marginally smaller for Seoul than for US cities and of order $\approx 5-8.10^4/$capita, and the typical value of housing estate is of order $\approx 10^4/m^2$), we  estimate that for each kilometer of highway, a $2$km wide stripe of land would see its value increase by $5.10^3/m^2$, leading to an external cost of $10^{10}/km$ associated to the presence of the highway. We represented this cost on the Fig.~\ref{fig:socialcost}, with the red area representing cities of population $>10^5$ inhabitants (i.e. which currently likely have a freeway) and for which the external costs of the freeway exceed its benefits. We observe that indeed for cities with population less than $1.4\times 10^6$ inhabitants, it is beneficial to remove the urban freeway. 

\section*{Discussion}

We showed empirically that US cities display two population thresholds at approximately $10^5$ and $10^6$ inhabitants above which urban freeways and ring roads respectively become generalized elements of the road infrastructure. We proposed a simple model to explain these transitions, based on a cost-benefit analysis and which focuses on the balance between the cost of the infrastructure and the value of time spent in traffic. This simple analysis provides results in good agreement with our empirical observations, in particular the existence of abrupt transitions in these systems. Incidentally, it is interesting to note that optimal networks also display such an abrupt transition \cite{Aldous:2019}. The cost-benefit analysis allows us to express the population thresholds in terms of relevant parameters such as the construction cost per unit length, the velocity on the ringroad, the value of time, and the capacity of roads. 

This framework also allows us to show that the negative externalities of urban freeways can lead in many cases to their removal. External costs are hard to quantify, and more importantly will vary strongly depending on the city (typically its attractivity and its population density). Our results should be used as a guideline to help taking this decision. Each city should evaluate the costs associated to the freeway and compare them to the value $\epsilon_c$ described in this paper. If the costs exceed this value, it means that it is not only detrimental to the environment but even economically unjustified to rebuild urban freeways. Realistically, the cost will be the sum of a large number of factors.

\begin{figure}[ht!]
	\includegraphics[width=.45\textwidth]{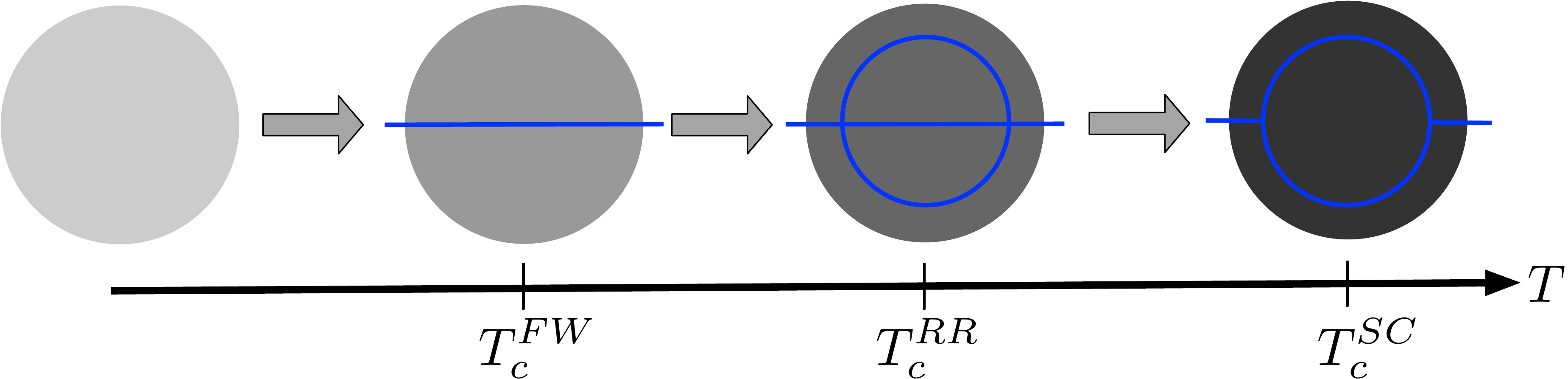}
	\caption{Summary of the evolution of urban infrastructure when the number of commuters $T$ grows. In the 1960s, urban freeways were built through city centers, followed shortly after by ring roads around the cities. Taking into account the detrimental aspects of urban freeways,  cities tend to remove their urban freeways.}
	\label{fig:social}
\end{figure}

The cost-benefit analysis thus provides an interesting and simple tool for assessing different scenarios of the evolution of infrastructures in large cities. Based on this analysis, we thus find that when the population and the number of commuters $T$ grow, the typical evolution scenario for road infrastructures can be summarized as shown in Fig.~\ref{fig:social}.

Our analysis is simplified and didn't take into account some effects such as induced demand \cite{Hymel:2010} \cite{Goodwin:1996}. Building new roads increase the traffic capacity and usually induce an increase in demand (e.g. by making it more attractive to live further away from the city center and travelling a larger distance). We expect the decision to remove urban freeways to be accompanied by a similar induced reduction of traffic demand, meaning that the actual threshold $\epsilon_c'$ for the social cost should actually probably be lower.  In fact, the few examples of cities which have already removed their urban freeways (e.g. Seoul, Paris...) tend to show that this reduction in traffic occurs \cite{Billings:2013}. The decision to remove a urban freeway should be accompanied by other measures regarding the infrastructure, which our model doesn't account for. Successful examples of freeway removals show that the freeway could be replaced by at-grade Boulevards  with high integration for pedestrians, vegetation and alternative forms of mobility, while the public transport offer should be developed to further help to induce a reduction of traffic demand \cite{Kraft-Klehm:2015} \cite{Billings:2013}. With our definition of urban freeways, we looked exclusively at highways crossing the city, but most US cities (even small), have the typical `stroad' \cite{Wiki:stroad} type of Boulevard, i.e. multilane relatively high capacity roads with at-grade intersections and traffic lights, which might be a little less intrusive in terms of urbanism, but still pose problems in terms of segregation and impracticality for pedestrians. We are not advocating for the replacement of urban freeways by this type of roads, as the social cost associated to them would still be important. Generally speaking, the discussion proposed here should serve as a guide for the decision process regarding freeway removals, which should be part of a more global reflection about the shape of our cities and the importance of cars. This is particularly important at a time when the majority of the infrastructure in the US needs to be rebuilt in the coming years or decade.

\section*{Acknowledgments}
ET thanks the IPhT for the financial support of his PhD thesis.

\bibliographystyle{prsty}

\onecolumngrid
\newpage
\begin{table}
\begin{tabular}{ | m{3.5cm} | m{10cm}| m{3.5cm} | }
  \hline
  \textbf{Variable} & \textbf{Description} & \textbf{Unit and value (if fixed)} \\   \hline
  
   \textbf{Population and traffic} &&\\ \hline
  $P$ & Population of the city & \\ 
  \hline
  $T$ & Number of car commuters & \\   \hline
  $h$ & Duration of the rush hours & $h=2$ hours \\   \hline
  $N$ & Number of rush hour periods for or cost-benefit analysis & $N=730$ \\ \hline

  \textbf{BPR function} && \\ \hline
  $\tau$ & Travel time on a road segment & hours \\ \hline
  $Q$ & Traffic on a road (flux) & cars/h \\ \hline
  $Q_c$ & Critical traffic, capacity of the road & $Q_c = 10^3$ cars/h \\ \hline
  $\mu$ & Critical exponent of the BPR & We suppose $\mu=1$ \\ \hline

  \textbf{Geometry} &&\\ \hline
  $R$ & Radius of the city & km \\ \hline
  $L, (2R-L)$ & Start and end of the freeway & km \\ \hline
  $\ell, (2R-\ell)$ & Start and end of the ringroad & km \\ \hline
  $g $ & Geometric factor for the length of ringroad & $g=\frac{\pi}{2}$ \\ \hline
  $v_0$ & Free-flow speed on urban streets & $v_0=20$ km/h \\ \hline
  $v$ & Free-flow speed on highways & $v=100$ km/h \\ \hline
  $\alpha$ & Part of commuters travelling to the city center & Typically $\alpha=50\%$ \\ \hline
  $\beta = 1-\alpha$ & Part of commuters travelling to other suburbs & Typically $\beta=50\%$ \\ \hline

 \textbf{Cost analysis} &&\\ \hline
  $\Delta \tau$ & Travel time variation between scenarii & hours \\ \hline
  $C$ & Cost associated to the travel time & $\$$ \\ \hline
  $\Delta C$ & Cost variation between scenarii & $\$$ \\ \hline
  $\epsilon$ & Cost of construction of a highway &  $\epsilon=10^7$ \$/km \\ \hline
  $V$ & Value of time & $V=20\$$/hour \\ \hline
  $T_c$ & Critical number of commuters leading to a change of infrastructure & \\ \hline

  \textbf{Compound variables} &&\\ \hline
  $x = \frac{L}{R}$ & dimensionless position of the freeway & \\ \hline
  $y = \frac{\ell}{R}$ & dimensionless position of the ringroad & \\ \hline
  $\eta = \frac{v_0}{v}$ & ratio between free-flow speeds & $\eta = 0.2$ \\ \hline
  $c = \frac{T}{P}$ & Fraction of the population who commutes by car & $c=40\%$ \\ \hline
 \label{table}
\end{tabular}

\caption{Definition of the variables used in this work.}
\end{table}

\end{document}